%% file: first-repro.tex
\pdfoutput=1
\documentclass{article}

\usepackage[final]{neurips_2024}

\usepackage[utf8]{inputenc} 
\usepackage[T1]{fontenc}    
\usepackage[hidelinks]{hyperref} 
\usepackage{url}            
\usepackage{booktabs}       
\usepackage{amsfonts}       
\usepackage{nicefrac}       
\usepackage{microtype}      
\usepackage{xcolor}         
\usepackage[hang,flushmargin]{footmisc}
\usepackage{enumitem}
\usepackage{xspace}
\usepackage{arydshln}
\usepackage{multirow}
\usepackage{graphicx}
\usepackage{pifont}
\usepackage{tikz}
\usepackage{subcaption}
\usepackage{natbib}
\usepackage{array}

\usepackage{etoolbox}
\makeatletter
\patchcmd{\@verbatim}
  {\verbatim@font}
  {\verbatim@font\small}
  {}{}
\makeatother

\newcommand{\ignore}[1]{}

\newcommand{\tableformat}{\fontsize{9pt}{12pt}\selectfont
\setlength{\tabcolsep}{3pt}
\renewcommand{\arraystretch}{1}}

\newcommand{\zb}{Zephyr$_\beta$\xspace}
\newcommand{\rz}{RankZephyr\xspace}
\newcommand{\rv}{RankVicuna\xspace}
\newcommand{\mistral}{Mistral-7B-Instruct-v0.3\xspace}
\newcommand{\llama}{LLaMA-3.1-8B-Instruct\xspace}
\newcommand{\reddy}{FIRST-Reddy\xspace}

\newcommand{\rfirstl}[1]{First\-#1\xspace}

\newcommand{\rflz}[0]{\rfirstl{RankZephyr}}
\newcommand{\rflzb}[0]{\rfirstl{Zephyr$_\beta$}}
\newcommand{\rfll}[0]{\rfirstl{LLaMA}}
\newcommand{\rflm}[0]{\rfirstl{Mistral}}
\newcommand{\rlm}[0]{RankMistral\xspace}

\newcommand{\likelihood}{\textsubscript{\emph{l}}\xspace}
\newcommand{\generation}{\textsubscript{\emph{g}}\xspace}

\newcommand{\rzl}{RankZephyr\likelihood}
\newcommand{\rzg}{RankZephyr\generation}
\newcommand{\rvl}{RankVicuna\likelihood}
\newcommand{\rvg}{RankVicuna\generation}

\newcommand{\rlml}{RankMistral\likelihood}
\newcommand{\rlmg}{RankMistral\generation}

\title{An Early FIRST Reproduction and Improvements to Single-Token Decoding for Fast Listwise Reranking}

\author{Zijian Chen, Ronak Pradeep, Jimmy Lin\\[1ex]
David R.\ Cheriton School of Computer Science\\
University of Waterloo
}

\begin{document}
\maketitle

\begin{abstract}
Recent advances have demonstrated that large language models (LLMs) excel as listwise rerankers, but their high computational demands remain a barrier to widespread adoption.
Further, the traditional language modeling (LM) objective is not ideally suited for reranking tasks.
FIRST is a novel approach that addresses these challenges by integrating a learning-to-rank objective and leveraging the logits of only the first generated token, thereby significantly reducing inference latency compared to traditional LLM rerankers.
In this study, we extend the evaluation of FIRST to the TREC Deep Learning datasets (DL19--22), validating its robustness across diverse domains.
We investigate the influence of different first-stage retrievers on FIRST rerankers, observing diminishing returns and patterns consistent with traditional LLM rerankers.
Through applying the FIRST objective to a broader range of backbone models, we achieve effectiveness surpassing the original implementation.
Our experiments confirm that fast reranking with single-token logits does not compromise out-of-domain reranking quality.
To better quantify the computational savings in the original study, we measure and compare latency to find a 21\%--42\% gain across various models and benchmarks.
Moreover, while LM training implicitly improves zero-shot single-token reranking, our experiments also raise questions about whether LM pre-training may hinder subsequent fine-tuning with the FIRST objective.
These findings pave the way for more efficient and effective listwise reranking in future applications.
\end{abstract}

\section{Introduction}

Large language models (LLMs) have emerged as powerful tools for information retrieval, including the task of document reranking \citep{zhu24:llm_ir_survey}.
Recent studies have demonstrated that LLMs, when employed as zero-shot listwise rerankers, can surpass traditional supervised approaches without requiring extensive relevance judgments \citep{sun23:rankgpt, ma23:listwise, unoduolisto}.

However, associated with LLM rerankers is their high inference latency, presenting challenges for practical deployment.
Traditional listwise reranking approaches frame the reranking problem as a generation task, requiring LLMs to produce complete permutations of the document identifiers as the output ranking, trained using language modeling loss against the correct permutation.
The combination of auto-regressive generation in transformers and the substantial size of LLMs leads to concerning latency issues in practical applications \citep{zhu24:llm_ir_survey}.

To address these efficiency challenges, \citet{zhuang24:setwise} proposed examining the relative magnitudes of first-token logits for reranking, eliminating the need to generate entire identifier permutations.
Further, recent work by \citet{reddy24:first} introduced FIRST (\textbf{F}aster \textbf{I}mproved Listwise \textbf{R}eranking with \textbf{S}ingle \textbf{T}oken Decoding), which not only leverages single-token reranking, but also combines it with a new training objective.
The authors claimed that the language modeling objective is fundamentally suboptimal for ranking as they uniformly penalize incorrect ranking across all positions, failing to emphasize the importance of correctly ranking the most relevant documents.
By incorporating a learning-to-rank loss alongside their single-token strategy, FIRST promised more efficient reranking without compromising effectiveness.
In this work, we present a comprehensive reproduction and analysis of the FIRST approach across multiple dimensions.
We extended its evaluation to the TREC Deep Learning Track datasets to test its robustness, investigated its interaction with various first-stage retrievers, and expanded its application to a broader range of backbone models, surpassing the original implementation in effectiveness.

Our experiments confirmed FIRST as an efficient alternative to traditional listwise LLM reranking, demonstrating latency improvements of 21\%--42\% across various models and benchmarks while maintaining effectiveness.
Moreover, our investigation yielded several insights about the relationship between language modeling and the effectiveness of FIRST: while language modeling training implicitly improves zero-shot single-token reranking capabilities, we discovered that language modeling pre-training may paradoxically hinder subsequent FIRST fine-tuning.

\section{Background and Related Work}

\paragraph{Multi-stage Ranking.}
Given a corpus of documents $\mathcal{C} = \{d_1, d_2, \cdots, d_n\}$ and a query $q$, \emph{retrieving} refers to the task of finding an ordered list $\mathcal{R}$ of $k$ most relevant documents from $\mathcal{C}$, in descending relevance with respect to $q$, where $k << |\mathcal{C}|$.
\emph{Reranking} refers to the downstream task that reorders $\mathcal{R}$, if necessary, to a more accurate ranking.
This retrieve-rerank procedure common in modern systems is referred to as \emph{multi-stage ranking}, where the retrieve step often uses a more computationally efficient approach, followed by a more accurate but expensive rerank step \citep{nogueira19:monobert}.
The retrieving system is referred to as a \emph{first-stage retriever} in this setting, while the reranking system is referred to as a \emph{reranker}.

\paragraph{Reranking with Large Language Models.}
Recent advancements have demonstrated that LLMs can serve as effective rerankers \citep{pradeep23:rankzephyr, qin24:pairwise}.
Such approaches can be categorized into pointwise, pairwise, or listwise.

Earlier works predominantly utilized the \emph{pointwise} approach, where the LLM assesses each query-document pair independently, computing a likelihood or binary relevance judgment in isolation \citep{zhuang24:binary}.
On the other hand, \emph{pairwise} approaches leverage the LLM to compare the relevance of two documents at a time, given the same query \citep{qin24:pairwise}.

More recently, RankGPT \citep{sun23:rankgpt} experimented with a \emph{listwise} approach, where the LLM is prompted with a query and a list of documents, generating a complete permutation of the documents based on their relevance to the query.
Subsequently, \rz by \citet{pradeep23:rankzephyr} continued this theme by instruction-tuning \zb \citep{tunstall23:zephyr} to perform such listwise document reranking.
Concurrently, Rank-without-GPT~\citep{zhang23:rankwogpt} explored instruction tuning leveraging non-GPT teacher models.
Newer work~\citep{tamber2023scalingdownlittingup} has also explored more efficient listwise reranking through smaller encoder-decoder models.

In this study, we focus on utilizing LLMs as listwise rerankers.

\paragraph{Listwise Reranking with FIRST.}
Although listwise reranking approaches have demonstrated strong effectiveness, they are computationally intensive, as they typically rely on sequence generation to produce an entire permutation of document identifiers.
FIRST by \citet{reddy24:first} addresses these inefficiencies by utilizing only the logits from the first token in the output sequence to determine the rank order of candidate documents, rather than generating a complete ranked sequence.

Further, FIRST incorporates a learning-to-rank objective during training, prioritizing ranking accuracy for top candidates over less relevant ones, rather than focusing solely on the language modeling objective.
This modification by \citeauthor{reddy24:first} promised to give more effective supervision during training.

Together, FIRST offers a faster, more efficient reranking model that achieves comparable or superior ranking quality with significantly reduced computational demands.
In this work, we refer to LLM rerankers that employ full sequence generation, such as \rz \citep{pradeep23:rankzephyr}, as ``traditional LLM rerankers'', while we designate models utilizing the single-token approach as ``FIRST rerankers''.

\section{Methods}

We begin by summarizing the methodology introduced by \citet{reddy24:first}, while also formalizing the problem and notation.

\paragraph{Listwise Reranking using Sliding Windows.}
Recall the reranking problem: given a candidate document list $\mathcal{R} = \{d_1, d_2, \dots, d_n\}$ for a query $q$, reorder $\mathcal{R}$ based on the document relevance to $q$.
Due to context window constraints in LLMs, this reordering typically cannot be accomplished in a single step.
Following \citet{sun23:rankgpt}, we employ a sliding window approach with window size $m$ and step size $s$.
The window processes $m$ documents at a time, moving from the end of the list toward the front with a stride of $s$ documents.
At each step, the LLM reorders the documents within the current window according to their relevance to $q$.

\paragraph{FIRST Objective.}
Conventional listwise reranking approaches require LLMs to generate a complete permutation of the candidate documents and are trained using a language modeling objective against the correct permutation.
In contrast, FIRST derives the rank ordering solely from the relative magnitude of the output logits of the first generated identifier token, eliminating the need for full sequence generation.
To formally state FIRST's objective:
\begin{itemize}
    \item $t_i$ denotes the identifier token for document $d_i$
    \item $p_i$ represents the logits for generating $t_i$ as the first identifier token
    \item $r_i \in \{1, \dots, m\}$ indicates the true rank of document $d_i$ among $m$ candidates
\end{itemize}
FIRST incorporates a weighted pairwise learning-to-rank loss defined as:
$$
\mathcal{L}_{Rank} = \sum_{r_i < r_j} \frac{1}{i + j} \log(1 + \exp(p_i - p_j))
$$
where the weight term $\frac{1}{i + j}$ prioritizes accurate ranking of more relevant documents.

The final training objective combines this ranking loss with the traditional language modeling loss $\mathcal{L}_{LM}$:
$\mathcal{L}_{Joint} = \mathcal{L}_{LM} + \lambda \mathcal{L}_{Rank}
$
where $\lambda$ is a hyperparameter set to 10 in the original work.
We will refer to this as the ``FIRST objective''.

\section{Experimental Setup}

\subsection{Reproducing FIRST}
\label{setup:repro}

First, we reproduced the results from \citet{reddy24:first} using the procedure detailed in their work.

\paragraph{Model and Training.} We initialized from \zb \citep{tunstall23:zephyr}, a 7B LLM instruction-tuned from Mistral-7B-Instruct-v0.1 \citep{jiang23:mistral7b} on chat datasets, and we fine-tuned using the joint objective $\mathcal{L}_{Joint}$ with $\lambda = 10$.
Training lasted for 3 epochs, using:
\begin{itemize}
    \item Effective batch size: 32
    \item Learning rate: 5e-6
    \item Noisy embeddings \citep{jain23:noise}
    \item Sliding window size ($m$): 20
    \item Step size ($s$): 10
\end{itemize}
All model training were performed on 4 NVIDIA RTX A6000's.
The trained checkpoint is referred to as \rflzb.

\paragraph{Training Data.} We utilized the same dataset as \citet{reddy24:first}: 40K GPT-4 labeled rerank instances from \citet{pradeep23:rankzephyr}.
Note that to ensure single-token identification, \citet{reddy24:first} has converted the original dataset to use alphabetical identifiers rather than numerical identifiers.\footnote{The alphabetical version is available at: \href{https://huggingface.co/datasets/rryisthebest/rank_zephyr_training_data_alpha}{https://huggingface.co/datasets/rryisthebest/rank\_zephyr\_training\_data\_alpha}}

\paragraph{Retriever.} Following the original paper, we employed Contriever \citep{lei23:contriever} as our first-stage retriever, selecting the top 100 documents for LLM reranking.

\paragraph{Baseline Evaluation.} We evaluated on the same data used in the original FIRST study, which comprises of several subsets of BEIR \citep{thakur21:beir} and MS MARCO \citep{bajaj18:msmarco}.

\subsection{Varying the Model Backbone}
\label{setup:backbones}

To validate the FIRST objective across different architectures, we fine-tuned various other prominent LLMs of comparable size, besides the original Zephyr$_\beta$, while keeping other settings constant:
\begin{itemize}
    \item \rflm: fine-tuned from \mistral \footnote{\href{https://huggingface.co/mistralai/Mistral-7B-Instruct-v0.3}{https://huggingface.co/mistralai/Mistral-7B-Instruct-v0.3}}
    \item \rfll : fine-tuned from \llama \footnote{\href{https://ai.meta.com/blog/meta-llama-3-1/}{https://ai.meta.com/blog/meta-llama-3-1}}
    \item \rflz : fine-tuned from \rz \citep{pradeep23:rankzephyr}
\end{itemize}
Note that \rflz was fine-tuned from \rz, which in turn has been previously trained as a listwise reranker on $\mathcal{L}_{LM}$.

\subsection{Evaluating on TREC Deep Learning}

To further analyze the robustness of FIRST across diverse domains, we extended the evaluation beyond the datasets selected by \citet{reddy24:first} to include 4 TREC Deep Learning Track test collections:

\paragraph{MS MARCO v1-based Collections:}
\begin{itemize}
    \item TREC 2019 Deep Learning Track (DL19) \citep{craswell19:dl19}
    \item TREC 2020 Deep Learning Track (DL20) \citep{craswell20:dl20}
\end{itemize}
These collections draw from the MS MARCO v1 passage corpus, containing approximately 8.8 million passages.
\paragraph{MS MARCO v2-based Collections:}
\begin{itemize}
    \item TREC 2021 Deep Learning Track (DL21) \citep{craswell21:dl21}
    \item TREC 2022 Deep Learning Track (DL22) \citep{craswell22:dl22}
\end{itemize}
These collections utilize the substantially larger MS MARCO v2 passage corpus, containing around 138 million passages.

\subsection{Evaluating Zero-Shot FIRST}

A key insight from \citet{reddy24:first} was that \rz, despite being trained solely with a language modeling objective ($\mathcal{L}_{LM}$), exhibited zero-shot ability to rank documents using the logits of the first identifier only; its capability to rerank using the FIRST strategy was notably stronger than in the pre-trained model, suggesting that fine-tuning on $\mathcal{L}_{LM}$ implicitly improves single-token ranking.
To investigate this phenomenon further, we evaluated two open-source listwise rerankers trained exclusively on $\mathcal{L}_{LM}$---\rz and \rv~\citep{pradeep23:rankvicuna, pradeep23:rankzephyr}---on the TREC Deep Learning Track test collections (TREC DL19--22), using only the first-token logits at inference time.

\subsection{Evaluating Different First-Stage Retrievers}

To evaluate the robustness of FIRST across the different first-stage retrievers, we conducted experiments with 3 first-stage retrievers, in combination with the models discussed in Section~\ref{setup:backbones}, on TREC DL19--22.
These retrievers represent diverse approaches to information retrieval: a traditional lexical method BM25~\citep{robertson09:bm25}, a sparse neural retriever SPLADE++ EnsembleDistil~\citep{formal22:spladepp}, and a dense neural retriever RepLLaMA~\citep{ma23:repllama}---all of which were retrievers employed in the \rz study~\citep{pradeep23:rankzephyr}.

\subsection{Comparing FIRST and Language Modeling}

To rigorously assess the effectiveness of the FIRST objective against the traditional language modeling objective $\mathcal{L}_{LM}$, we conducted ablation experiments on two pre-trained models.
First, we compared \rflzb and \rz, both initialized from \zb, but \rflzb was fine-tuned on FIRST while \rz was fine-tuned on $\mathcal{L}_{LM}$.
Similarly, we evaluated \rflm against \rlm, both initialized from \mistral but \rflm on FIRST and \rlm on $\mathcal{L}_{LM}$.
Aside from the base pre-trained model, all other settings remain the same as in Section~\ref{setup:repro}.

\subsection{Analyzing Latency}

To quantify the computational efficiency gains of FIRST's single-token approach versus full sequence generation, we measured inference latency using NVIDIA Nsight Systems\footnote{\href{https://developer.nvidia.com/nsight-systems}{https://developer.nvidia.com/nsight-systems}} across the TREC DL19--22 datasets.
All latency experiments were conducted on a single NVIDIA RTX A6000 to reduce confounding factors stemming from orchestration across multiple GPUs.

\section{Results and Discussion}

\subsection{Reproducing FIRST}

\input{tables/beir_eval}

Table~\ref{tab:beir_eval} compares the reranking quality across various models trained with the FIRST objective, evaluated on the benchmark datasets used in the original study.
Following the original experimental setup, we employed Contriever \citep{lei23:contriever} as the first-stage retriever, retrieving the top 100 documents.
We compared two key implementations: \reddy, the official checkpoint\footnote{\href{https://huggingface.co/rryisthebest/First_Model}{https://huggingface.co/rryisthebest/First\_Model}} released by \citeauthor{reddy24:first} ran on our machines, and \rflzb, our reproduction trained according to \citeauthor{reddy24:first}'s procedures detailed in Section~\ref{setup:repro}.

Our reproduced model achieved comparable average effectiveness to the original checkpoint, while demonstrating modest improvements across most datasets.
This validates both our implementation and the reproducibility of the FIRST approach.

\subsection{Varying the Model Backbone}

As demonstrated in Table~\ref{tab:beir_eval}, \rflm achieved the highest nDCG@10 scores on 8 out of 11 of the datasets selected by \citet{reddy24:first}.
This superior effectiveness can likely be attributed to its initialization from a more recent version of Mistral 7B, while \reddy, \rflzb, and \rflz were fine-tuned from \zb, which in turn was fine-tuned from an earlier Mistral 7B version.

\begin{figure}[t]
    \centering
    \includegraphics[width=\textwidth]{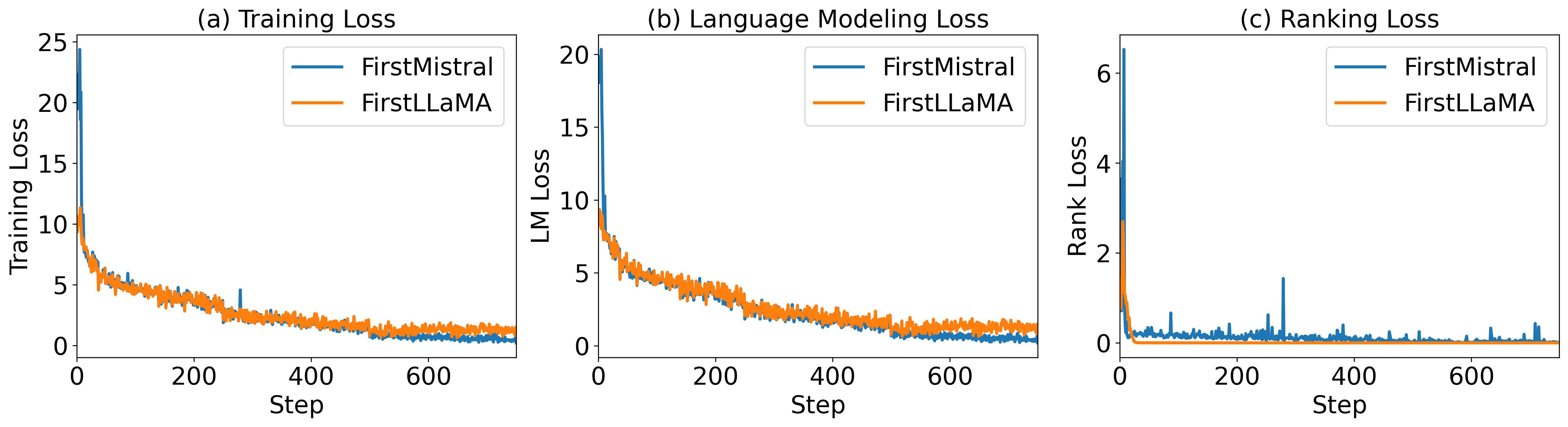}
    \caption{Comparison of training, language modeling, and ranking losses for \rflm and \rfll during training.}
    \label{fig:loss_comparison}
\end{figure}

To better understand \rfll's relatively lower effectiveness, we tracked and compared its training process with \rflm, as illustrated in Figure~\ref{fig:loss_comparison}.
The analysis monitored three metrics: the combined training loss ($\mathcal{L}_{Joint}$), its language modeling component ($\mathcal{L}_{LM}$), and its ranking component ($\lambda \mathcal{L}_{Rank}$).
While \rfll exhibited faster convergence in ranking loss, \rflm demonstrated more efficient convergence in language modeling loss, which dominated the overall training loss.

\subsection{Evaluating on TREC Deep Learning}

\input{tables/dl_eval}

Table~\ref{tab:dl_eval} presents a comprehensive evaluation of model effectiveness across TREC DL19--22.
The results reinforced our previous findings from Sections 5.1 and 5.2: \rflm maintained its superior scores across these new test collections, while \reddy and \rflzb demonstrated comparable effectiveness, and \rfll continued to show relatively lower effectiveness.

Notably, with the exception of \rfll, models trained with the FIRST objective achieved effectiveness comparable to \rzg, with \rflm even surpassing \rzg's average effectiveness.
Since TREC DL19--22 were not included in the original study, these results provide additional validation for \citeauthor{reddy24:first}'s central claim that the FIRST approach, their attempt to efficient reranking, does not compromise reranking effectiveness.

\subsection{Evaluating Zero-Shot FIRST}

As evident in row \rzl of Table~\ref{tab:dl_eval}, \rz achieved effectiveness on par with dedicated FIRST models when using only first-token logits for reranking, despite not being explicitly trained for this objective.
This observation validates \citeauthor{reddy24:first}'s hypothesis that training on $\mathcal{L}_{LM}$ implicitly improves the model's ability to perform single-token rerank.
In fact, this is even more pronounced in the case of \rv, where \rvl outperformed \rvg; that is, a model trained on $\mathcal{L}_{LM}$ only was even more effective on average when reranking using a single-token.

However, the relationship between $\mathcal{L}_{LM}$ and FIRST is slightly more nuanced.
We compared two models: 
\begin{enumerate}
    \item [1.] \rflz: sequentially fine-tuned first on $\mathcal{L}_{LM}$ (starting from \zb to create \rz) and then on the FIRST objective
    \item [2.] \rflzb: fine-tuned directly from \zb using the FIRST objective
\end{enumerate}
The results in Tables~\ref{tab:beir_eval} and~\ref{tab:dl_eval} show that \rflzb consistently outperformed \rflz across most datasets.
That is, while $\mathcal{L}_{LM}$ training improves zero-shot FIRST effectiveness, it may actually hinder subsequent fine-tuning with the FIRST objective.
This result challenges the intuitive assumption that language model pre-training necessarily benefits downstream FIRST training.

\subsection{Evaluating Different First-Stage Retrievers}

\input{tables/first-stage}

Table~\ref{tab:retrievers} presents an analysis of different first-stage retrievers paired with \rflm, our most effective model.
The results revealed two key patterns.
First, stronger initial retrieval effectiveness consistently lead to better post-reranking effectiveness.
Second, we observed diminishing  returns from better first-stage retrievers, as evidenced by the decreasing percentage improvements shown in Table~\ref{tab:retrievers}.

These findings align with patterns reported in \rz \citep{pradeep23:rankzephyr}, suggesting that the relationship between first-stage retriever quality and final ranking effectiveness remains consistent, regardless of whether the reranker employs traditional listwise reranking with full generation or FIRST's single-token approach.

\subsection{Comparing FIRST and Language Modeling}

\input{tables/lm_mistral}

From Table~\ref{tab:dl_eval}, we see that \rflzb and \rz demonstrated comparable effectiveness, with \rz showing a slight edge on average.
To further investigate the effectiveness of $\mathcal{L}_{LM}$ vs.
FIRST, we conducted additional experiments comparing \rflm and \rlm, as shown in Table~\ref{tab:lm_mistral}.
The consistent effectiveness in both models across all datasets in this comparison provides robust evidence that, despite using only the first token, the FIRST objective maintains its ranking effectiveness.

\subsection{Analyzing Latency}

\input{tables/wall_clock_time_dl}

Table~\ref{tab:latency} compares the inference latency between full sequence generation and FIRST's single-token approach across 3 models: \rz, \rv, and \rlm.
The results demonstrated substantial computational savings, with FIRST reducing inference time by 21\% to 42\% across the 3 models.
These efficiency gains, consistent across different model architectures, confirm that FIRST is a competitive listwise reranking approach, offering faster inference while maintaining the effectiveness demonstrated in previous sections.
Such latency benefits are particularly critical as these models are increasingly deployed in real-world serving settings, where response time directly impacts user experience and infrastructure costs~\citep{ragnarok}.

\section{Conclusion}

Through a comprehensive study extending FIRST across evaluation datasets, first-stage retrievers, and backbone models, our study expanded understanding of FIRST's capabilities, validating its promise as a more efficient yet effective alternative to traditional LLM reranking.

Beyond confirming FIRST's reproducibility, robustness, and performance, our investigation revealed several key insights.
We found that different first-stage retrievers exhibit patterns of diminishing returns when combined with FIRST, mirroring observations from traditional LLM rerankers.
This suggests that despite changes in both inference approach and training objective, the fundamental dynamics of multi-stage retrieval remain consistent.

Further, our study demonstrated that while LM training does improve zero-shot single-token reranking capabilities, it may paradoxically hinder subsequent FIRST fine-tuning.
This counterintuitive result raises questions about the relationship between the language modeling and ranking objectives.
Moreover, \rfll's relatively poor effectiveness, combined with its rapid convergence to ranking loss but slow convergence to generation loss during training, suggests directions for future research to explore alternative learning-to-rank losses in combination with single-token reranking.

\section*{Acknowledgments}

This research was supported in part by the Natural Sciences and Engineering Research Council (NSERC) of Canada.
Additional funding is provided by Microsoft via the Accelerating Foundation Models Research program.

\bibliographystyle{ACM-Reference-Format}
\bibliography{first-repro}

\end{document}

%% file: tables/beir_eval.tex

\begin{table*}
\begin{center}
\tableformat
    \begin{tabular}{cccccc}
    \toprule
    \textbf{Dataset} & \textbf{\reddy} & \textbf{\rflzb} & \textbf{\rflz} & \textbf{\rflm} & \textbf{\rfll} \\
    \midrule
    climate-fever & \textbf{0.2672} & 0.2314 & \underline{0.2519} & 0.2417 & 0.2434 \\
    dbpedia-entity & \textbf{0.5092} & 0.4908 & 0.4761 & \underline{0.5033} & 0.4387 \\
    fever & 0.8164 & 0.8215 & 0.7927 & \textbf{0.8413} & \underline{0.8301} \\
    fiqa & 0.4223 & \underline{0.4509} & 0.4263 & \textbf{0.4778} & 0.4162 \\
    hotpotqa & 0.7424 & \underline{0.7620} & 0.7506 & \textbf{0.7705} & 0.7017 \\
    msmarco & \underline{0.4425} & 0.4383 & 0.4275 & \textbf{0.4512} & 0.4316 \\
    nfcorpus & 0.3725 & \underline{0.3729} & 0.3555 & \textbf{0.3816} & 0.3481 \\
    nq & 0.6638 & \underline{0.6928} & 0.6535 & \textbf{0.6985} & 0.6290 \\
    scidocs & 0.2047 & \underline{0.2064} & 0.1874 & \textbf{0.2110} & 0.1848 \\
    scifact & 0.7459 & \underline{0.7680} & 0.7495 & \textbf{0.7769} & 0.7489 \\
    trec-covid & \textbf{0.7913} & \underline{0.7683} & 0.7552 & 0.7666 & 0.6565 \\
    \midrule
    Average & 0.5435 & \underline{0.5458} & 0.5297 & \textbf{0.5564} & 0.5117 \\
    \bottomrule
    \end{tabular}
\end{center}
\caption{Comparison of nDCG@10 across the datasets selected by \citet{reddy24:first} from BEIR and MS MARCO, on models trained on different backbones with the FIRST objective. Contriever was used as the first-stage retriever \citep{lei23:contriever}.}
\label{tab:beir_eval}

\end{table*}


%% file: tables/dl_eval.tex
\begin{table*}
\begin{center}
\tableformat
\begin{tabular}{lccccc}
\toprule
\textbf{Model} & \textbf{DL19} & \textbf{DL20} & \textbf{DL21} & \textbf{DL22} & \textbf{Average} \\
\midrule

FIRST-Reddy & 0.7476 & \underline{0.7986} & \textbf{0.7709} & 
\underline{0.6944} & 0.7529 \\

\midrule

\rflzb & 0.7576 & 0.7550 & 0.7439 & 0.6767 & 0.7333 \\

\midrule

\rflz & 0.7315 & 0.7363 & 0.7145 & 0.6442 & 0.7066 \\

\midrule

\rflm & \underline{0.7678} & 0.7851 & \underline{0.7694} & \textbf{0.7030} & \textbf{0.7563} \\

\midrule

\rfll & 0.7363 & 0.7263 & 0.6941 & 0.6042 & 0.6902 \\

\midrule

\rzl & 0.7369 & 0.7370 & 0.7103 & 0.6165 & 0.7002 \\
\rzg & \textbf{0.7760} & \textbf{0.8140} & 0.7605 & 0.6669 & \underline{0.7543} \\

\midrule

\rvl & 0.7302 & 0.7103 & 0.6809 & 0.5794 & 0.6752 \\
\rvg & 0.6894 & 0.7081 & 0.6947 & 0.5521 & 0.6611 \\

\bottomrule
\end{tabular}
\end{center}
\caption{Comparison of nDCG@10 across TREC DL19--22 on models trained on the FIRST objective, as well as \rz and \rv that were trained on $\mathcal{L}_{LM}$. For \rz, \rzl denotes the model reranking using the logits of the first identifier only, and \rzg denotes the model reranking by generating the full permutation of document identifiers. \rvl and \rvg are defined similarly for \rv.}
\label{tab:dl_eval}
\end{table*}

%% file: tables/first-stage.tex
\begin{table*}
\centering
\tableformat
\begin{tabular}{lcc}
\toprule
\textbf{Method} & \textbf{DL19} & \textbf{DL20} \\

\midrule

BM25 & 0.5058 & 0.4796 \\
BM25 $\rightarrow$ \rflm & 0.7277 & 0.6971 \\
Improvement & +43.87\% & +45.35\% \\

\midrule

SPLADE++ EnsembleDistil & 0.7308 & 0.7197 \\ 
SPLADE++ EnsembleDistil $\rightarrow$ \rflm & 0.7678 & 0.7851 \\
Improvement & +5.06\% & +9.09\% \\

\midrule

RepLLaMA & 0.7384 & 0.7195 \\
RepLLaMA $\rightarrow$ \rflm & 0.7587 & 0.7682 \\
Improvement & +2.75\% & +6.77\% \\

\bottomrule
\end{tabular}
\caption{Comparison of nDCG@10 for \rflm across different first-stage retrievers, evaluated on DL19 and DL20.}
\label{tab:retrievers}
\end{table*}

%% file: tables/lm_mistral.tex
\begin{table*}

\begin{center}
\tableformat
    \begin{tabular}{cccccc}
    \toprule
    \textbf{Dataset} & \textbf{\rflm} & \textbf{\rlm} \\
    \midrule
    climate-fever & 0.2417 & 0.2411 \\
    dbpedia-entity & 0.5033 & 0.5088 \\
    fever & 0.8413 & 0.8223 \\
    fiqa & 0.4778 & 0.4537 \\
    hotpotqa & 0.7705 & 0.7349 \\
    msmarco & 0.4512 & 0.4351 \\
    nfcorpus & 0.3816 & 0.3828 \\
    nq & 0.6985 & 0.6835 \\
    scidocs & 0.2110 & 0.2108 \\
    scifact & 0.7769 & 0.7743 \\
    trec-covid & 0.7666 & 0.7840 \\
    \midrule
    DL19 & 0.7678 & 0.7772 \\
    DL20 & 0.7851 & 0.7949 \\
    DL21 & 0.7694 & 0.7603 \\
    DL22 & 0.7030 & 0.6980 \\
    \midrule
    Average & 0.6468 & 0.6407 \\
    \bottomrule
    \end{tabular}
\end{center}
\caption{Comparison of nDCG@10 for \rflm and \rlm across the datasets selected by Reddy et al., as well as TREC DL19--22.}
\label{tab:lm_mistral}
\end{table*}

%% file: tables/wall_clock_time_dl.tex
\begin{table*}
\centering
\tableformat
\begin{tabular}{lrrrr}
\toprule
\textbf{Model} & \textbf{DL19} & \textbf{DL20} & \textbf{DL21} & \textbf{DL22} \\
\midrule

\rzg & 142.3 & 174.3 & 147.8 & 211.3 \\
\rzl & 107.1 & 129.7 & 107.4 & 160.4 \\
Speedup & +24.7\% & +25.6\% & +27.3\% & +24.1\% \\
\midrule

\rvg & 218.4 & 261.7 & 216.8 & 320.0 \\
\rvl & 126.4 & 152.0 & 124.3 & 183.1 \\
Speedup & +42.1\% & +41.9\% & +42.7\% & +42.8\% \\
\midrule

\rlmg & 152.7 & 169.1 & 141.6 & 201.5 \\
\rlml & 102.4 & 128.2 & 104.8 & 157.7 \\
Speedup & +32.9\% & +24.2\% & +26.0\% & +21.7\% \\






\bottomrule
\end{tabular}
\caption{Comparison of wall clock time (s) across TREC DL19--22 on various models. For a model $M$, $M$\likelihood denotes the model reranking using the logits of the first identifier only, and $M$\generation denotes the model reranking by generating the full permutation of document identifiers.}
\label{tab:latency}
\end{table*}

%% file: first-repro.bbl

\begin{thebibliography}{27}


\ifx \showCODEN    \undefined \def \showCODEN     #1{\unskip}     \fi
\ifx \showDOI      \undefined \def \showDOI       #1{#1}\fi
\ifx \showISBNx    \undefined \def \showISBNx     #1{\unskip}     \fi
\ifx \showISBNxiii \undefined \def \showISBNxiii  #1{\unskip}     \fi
\ifx \showISSN     \undefined \def \showISSN      #1{\unskip}     \fi
\ifx \showLCCN     \undefined \def \showLCCN      #1{\unskip}     \fi
\ifx \shownote     \undefined \def \shownote      #1{#1}          \fi
\ifx \showarticletitle \undefined \def \showarticletitle #1{#1}   \fi
\ifx \showURL      \undefined \def \showURL       {\relax}        \fi
\providecommand\bibfield[2]{#2}
\providecommand\bibinfo[2]{#2}
\providecommand\natexlab[1]{#1}
\providecommand\showeprint[2][]{arXiv:#2}

\bibitem[\protect\citeauthoryear{Bajaj, Campos, Craswell, Deng, Gao, Liu,
  Majumder, McNamara, Mitra, Nguyen, Rosenberg, Song, Stoica, Tiwary, and
  Wang}{Bajaj et~al\mbox{.}}{2018}]%
        {bajaj18:msmarco}
\bibfield{author}{\bibinfo{person}{Payal Bajaj}, \bibinfo{person}{Daniel
  Campos}, \bibinfo{person}{Nick Craswell}, \bibinfo{person}{Li Deng},
  \bibinfo{person}{Jianfeng Gao}, \bibinfo{person}{Xiaodong Liu},
  \bibinfo{person}{Rangan Majumder}, \bibinfo{person}{Andrew McNamara},
  \bibinfo{person}{Bhaskar Mitra}, \bibinfo{person}{Tri Nguyen},
  \bibinfo{person}{Mir Rosenberg}, \bibinfo{person}{Xia Song},
  \bibinfo{person}{Alina Stoica}, \bibinfo{person}{Saurabh Tiwary}, {and}
  \bibinfo{person}{Tong Wang}.} \bibinfo{year}{2018}\natexlab{}.
\newblock \bibinfo{title}{MS MARCO: A Human Generated MAchine Reading
  COmprehension Dataset}.
\newblock
\newblock
\showeprint[arxiv]{1611.09268}~[cs.CL]
\urldef\tempurl%
\url{https://arxiv.org/abs/1611.09268}
\showURL{%
\tempurl}


\bibitem[\protect\citeauthoryear{Craswell, Mitra, Yilmaz, and Campos}{Craswell
  et~al\mbox{.}}{2020}]%
        {craswell20:dl20}
\bibfield{author}{\bibinfo{person}{Nick Craswell}, \bibinfo{person}{Bhaskar
  Mitra}, \bibinfo{person}{Emine Yilmaz}, {and} \bibinfo{person}{Daniel
  Campos}.} \bibinfo{year}{2020}\natexlab{}.
\newblock \showarticletitle{Overview of the {TREC} 2020 Deep Learning Track}.
  In \bibinfo{booktitle}{\emph{Proceedings of the Twenty-Ninth Text REtrieval
  Conference Proceedings (TREC 2020)}}. \bibinfo{address}{Gaithersburg,
  Maryland}.
\newblock


\bibitem[\protect\citeauthoryear{Craswell, Mitra, Yilmaz, Campos, and
  Lin}{Craswell et~al\mbox{.}}{2021}]%
        {craswell21:dl21}
\bibfield{author}{\bibinfo{person}{Nick Craswell}, \bibinfo{person}{Bhaskar
  Mitra}, \bibinfo{person}{Emine Yilmaz}, \bibinfo{person}{Daniel Campos},
  {and} \bibinfo{person}{Jimmy Lin}.} \bibinfo{year}{2021}\natexlab{}.
\newblock \showarticletitle{Overview of the {TREC} 2021 deep learning track}.
  In \bibinfo{booktitle}{\emph{Proceedings of the Thirtieth Text REtrieval
  Conference (TREC 2021)}}.
\newblock


\bibitem[\protect\citeauthoryear{Craswell, Mitra, Yilmaz, Campos, Lin,
  Voorhees, and Soboroff}{Craswell et~al\mbox{.}}{2022}]%
        {craswell22:dl22}
\bibfield{author}{\bibinfo{person}{Nick Craswell}, \bibinfo{person}{Bhaskar
  Mitra}, \bibinfo{person}{Emine Yilmaz}, \bibinfo{person}{Daniel Campos},
  \bibinfo{person}{Jimmy Lin}, \bibinfo{person}{Ellen~M. Voorhees}, {and}
  \bibinfo{person}{Ian Soboroff}.} \bibinfo{year}{2022}\natexlab{}.
\newblock \showarticletitle{Overview of the {TREC} 2022 Deep Learning Track}.
  In \bibinfo{booktitle}{\emph{Proceedings of the Thirty-First Text REtrieval
  Conference (TREC 2022)}}. \bibinfo{address}{Gaithersburg, Maryland}.
\newblock


\bibitem[\protect\citeauthoryear{Craswell, Mitra, Yilmaz, Campos, and
  Voorhees}{Craswell et~al\mbox{.}}{2019}]%
        {craswell19:dl19}
\bibfield{author}{\bibinfo{person}{Nick Craswell}, \bibinfo{person}{Bhaskar
  Mitra}, \bibinfo{person}{Emine Yilmaz}, \bibinfo{person}{Daniel Campos},
  {and} \bibinfo{person}{Ellen~M. Voorhees}.} \bibinfo{year}{2019}\natexlab{}.
\newblock \showarticletitle{Overview of the {TREC} 2019 Deep Learning Track}.
  In \bibinfo{booktitle}{\emph{Proceedings of the Twenty-Eighth Text REtrieval
  Conference Proceedings (TREC 2019)}}. \bibinfo{address}{Gaithersburg,
  Maryland}.
\newblock


\bibitem[\protect\citeauthoryear{Formal, Lassance, Piwowarski, and
  Clinchant}{Formal et~al\mbox{.}}{2022}]%
        {formal22:spladepp}
\bibfield{author}{\bibinfo{person}{Thibault Formal}, \bibinfo{person}{Carlos
  Lassance}, \bibinfo{person}{Benjamin Piwowarski}, {and}
  \bibinfo{person}{St\'{e}phane Clinchant}.} \bibinfo{year}{2022}\natexlab{}.
\newblock \showarticletitle{From Distillation to Hard Negative Sampling: Making
  Sparse Neural IR Models More Effective}. In
  \bibinfo{booktitle}{\emph{Proceedings of the 45th International ACM SIGIR
  Conference on Research and Development in Information Retrieval}} (Madrid,
  Spain) \emph{(\bibinfo{series}{SIGIR '22})}. \bibinfo{publisher}{Association
  for Computing Machinery}, \bibinfo{address}{New York, NY, USA},
  \bibinfo{pages}{2353–2359}.
\newblock
\showISBNx{9781450387323}
\urldef\tempurl%
\url{https://doi.org/10.1145/3477495.3531857}
\showDOI{\tempurl}


\bibitem[\protect\citeauthoryear{Jain, yeh Chiang, Wen, Kirchenbauer, Chu,
  Somepalli, Bartoldson, Kailkhura, Schwarzschild, Saha, Goldblum, Geiping, and
  Goldstein}{Jain et~al\mbox{.}}{2023}]%
        {jain23:noise}
\bibfield{author}{\bibinfo{person}{Neel Jain}, \bibinfo{person}{Ping yeh
  Chiang}, \bibinfo{person}{Yuxin Wen}, \bibinfo{person}{John Kirchenbauer},
  \bibinfo{person}{Hong-Min Chu}, \bibinfo{person}{Gowthami Somepalli},
  \bibinfo{person}{Brian~R. Bartoldson}, \bibinfo{person}{Bhavya Kailkhura},
  \bibinfo{person}{Avi Schwarzschild}, \bibinfo{person}{Aniruddha Saha},
  \bibinfo{person}{Micah Goldblum}, \bibinfo{person}{Jonas Geiping}, {and}
  \bibinfo{person}{Tom Goldstein}.} \bibinfo{year}{2023}\natexlab{}.
\newblock \bibinfo{title}{NEFTune: Noisy Embeddings Improve Instruction
  Finetuning}.
\newblock
\newblock
\showeprint[arxiv]{2310.05914}~[cs.CL]
\urldef\tempurl%
\url{https://arxiv.org/abs/2310.05914}
\showURL{%
\tempurl}


\bibitem[\protect\citeauthoryear{Jiang, Sablayrolles, Mensch, Bamford, Chaplot,
  de~las Casas, Bressand, Lengyel, Lample, Saulnier, Lavaud, Lachaux, Stock,
  Scao, Lavril, Wang, Lacroix, and Sayed}{Jiang et~al\mbox{.}}{2023}]%
        {jiang23:mistral7b}
\bibfield{author}{\bibinfo{person}{Albert~Q. Jiang}, \bibinfo{person}{Alexandre
  Sablayrolles}, \bibinfo{person}{Arthur Mensch}, \bibinfo{person}{Chris
  Bamford}, \bibinfo{person}{Devendra~Singh Chaplot}, \bibinfo{person}{Diego
  de~las Casas}, \bibinfo{person}{Florian Bressand}, \bibinfo{person}{Gianna
  Lengyel}, \bibinfo{person}{Guillaume Lample}, \bibinfo{person}{Lucile
  Saulnier}, \bibinfo{person}{Lélio~Renard Lavaud},
  \bibinfo{person}{Marie-Anne Lachaux}, \bibinfo{person}{Pierre Stock},
  \bibinfo{person}{Teven~Le Scao}, \bibinfo{person}{Thibaut Lavril},
  \bibinfo{person}{Thomas Wang}, \bibinfo{person}{Timothée Lacroix}, {and}
  \bibinfo{person}{William~El Sayed}.} \bibinfo{year}{2023}\natexlab{}.
\newblock \bibinfo{title}{Mistral 7B}.
\newblock
\newblock
\showeprint[arxiv]{2310.06825}~[cs.CL]
\urldef\tempurl%
\url{https://arxiv.org/abs/2310.06825}
\showURL{%
\tempurl}


\bibitem[\protect\citeauthoryear{Lassance, Pradeep, and Lin}{Lassance
  et~al\mbox{.}}{2024}]%
        {unoduolisto}
\bibfield{author}{\bibinfo{person}{Carlos Lassance}, \bibinfo{person}{Ronak
  Pradeep}, {and} \bibinfo{person}{Jimmy Lin}.}
  \bibinfo{year}{2024}\natexlab{}.
\newblock \showarticletitle{{Naverloo} @ {TREC} {Deep} {Learning} and {NeuCLIR}
  2023: As Easy as Zero, One, Two, Three — Cascading Dual Encoders, Mono,
  Duo, and Listo for Ad-Hoc Retrieval}. In
  \bibinfo{booktitle}{\emph{Proceedings of the Thirty-Second Text REtrieval
  Conference (TREC 2023)}}. NIST, \bibinfo{address}{Gaithersburg, Maryland}.
\newblock


\bibitem[\protect\citeauthoryear{Lei, Ding, Cao, Zan, Yates, and Tao}{Lei
  et~al\mbox{.}}{2023}]%
        {lei23:contriever}
\bibfield{author}{\bibinfo{person}{Yibin Lei}, \bibinfo{person}{Liang Ding},
  \bibinfo{person}{Yu Cao}, \bibinfo{person}{Changtong Zan},
  \bibinfo{person}{Andrew Yates}, {and} \bibinfo{person}{Dacheng Tao}.}
  \bibinfo{year}{2023}\natexlab{}.
\newblock \showarticletitle{Unsupervised Dense Retrieval with Relevance-Aware
  Contrastive Pre-Training}. In \bibinfo{booktitle}{\emph{Findings of the
  Association for Computational Linguistics: ACL 2023}},
  \bibfield{editor}{\bibinfo{person}{Anna Rogers}, \bibinfo{person}{Jordan
  Boyd-Graber}, {and} \bibinfo{person}{Naoaki Okazaki}} (Eds.).
  \bibinfo{publisher}{Association for Computational Linguistics},
  \bibinfo{address}{Toronto, Canada}, \bibinfo{pages}{10932--10940}.
\newblock
\urldef\tempurl%
\url{https://doi.org/10.18653/v1/2023.findings-acl.695}
\showDOI{\tempurl}


\bibitem[\protect\citeauthoryear{Ma, Wang, Yang, Wei, and Lin}{Ma
  et~al\mbox{.}}{2024}]%
        {ma23:repllama}
\bibfield{author}{\bibinfo{person}{Xueguang Ma}, \bibinfo{person}{Liang Wang},
  \bibinfo{person}{Nan Yang}, \bibinfo{person}{Furu Wei}, {and}
  \bibinfo{person}{Jimmy Lin}.} \bibinfo{year}{2024}\natexlab{}.
\newblock \showarticletitle{Fine-Tuning LLaMA for Multi-Stage Text Retrieval}.
  In \bibinfo{booktitle}{\emph{Proceedings of the 47th International ACM SIGIR
  Conference on Research and Development in Information Retrieval}} (Washington
  DC, USA) \emph{(\bibinfo{series}{SIGIR '24})}.
  \bibinfo{publisher}{Association for Computing Machinery},
  \bibinfo{address}{New York, NY, USA}, \bibinfo{pages}{2421–2425}.
\newblock
\showISBNx{9798400704314}
\urldef\tempurl%
\url{https://doi.org/10.1145/3626772.3657951}
\showDOI{\tempurl}


\bibitem[\protect\citeauthoryear{Ma, Zhang, Pradeep, and Lin}{Ma
  et~al\mbox{.}}{2023}]%
        {ma23:listwise}
\bibfield{author}{\bibinfo{person}{Xueguang Ma}, \bibinfo{person}{Xinyu Zhang},
  \bibinfo{person}{Ronak Pradeep}, {and} \bibinfo{person}{Jimmy Lin}.}
  \bibinfo{year}{2023}\natexlab{}.
\newblock \bibinfo{title}{Zero-Shot Listwise Document Reranking with a Large
  Language Model}.
\newblock
\newblock
\showeprint[arxiv]{2305.02156}~[cs.IR]
\urldef\tempurl%
\url{https://arxiv.org/abs/2305.02156}
\showURL{%
\tempurl}


\bibitem[\protect\citeauthoryear{Nogueira, Yang, Cho, and Lin}{Nogueira
  et~al\mbox{.}}{2019}]%
        {nogueira19:monobert}
\bibfield{author}{\bibinfo{person}{Rodrigo Nogueira}, \bibinfo{person}{Wei
  Yang}, \bibinfo{person}{Kyunghyun Cho}, {and} \bibinfo{person}{Jimmy Lin}.}
  \bibinfo{year}{2019}\natexlab{}.
\newblock \bibinfo{title}{Multi-Stage Document Ranking with BERT}.
\newblock
\newblock
\showeprint[arxiv]{1910.14424}~[cs.IR]
\urldef\tempurl%
\url{https://arxiv.org/abs/1910.14424}
\showURL{%
\tempurl}


\bibitem[\protect\citeauthoryear{Pradeep, Sharifymoghaddam, and Lin}{Pradeep
  et~al\mbox{.}}{2023a}]%
        {pradeep23:rankvicuna}
\bibfield{author}{\bibinfo{person}{Ronak Pradeep}, \bibinfo{person}{Sahel
  Sharifymoghaddam}, {and} \bibinfo{person}{Jimmy Lin}.}
  \bibinfo{year}{2023}\natexlab{a}.
\newblock \showarticletitle{{RankVicuna}: Zero-Shot Listwise Document Reranking
  with Open-Source Large Language Models}.
\newblock \bibinfo{journal}{\emph{arXiv:2309.15088}} (\bibinfo{year}{2023}).
\newblock


\bibitem[\protect\citeauthoryear{Pradeep, Sharifymoghaddam, and Lin}{Pradeep
  et~al\mbox{.}}{2023b}]%
        {pradeep23:rankzephyr}
\bibfield{author}{\bibinfo{person}{Ronak Pradeep}, \bibinfo{person}{Sahel
  Sharifymoghaddam}, {and} \bibinfo{person}{Jimmy Lin}.}
  \bibinfo{year}{2023}\natexlab{b}.
\newblock \bibinfo{title}{RankZephyr: Effective and Robust Zero-Shot Listwise
  Reranking is a Breeze!}
\newblock
\newblock
\showeprint[arxiv]{2312.02724}~[cs.IR]
\urldef\tempurl%
\url{https://arxiv.org/abs/2312.02724}
\showURL{%
\tempurl}


\bibitem[\protect\citeauthoryear{Pradeep, Thakur, Sharifymoghaddam, Zhang,
  Nguyen, Campos, Craswell, and Lin}{Pradeep et~al\mbox{.}}{2024}]%
        {ragnarok}
\bibfield{author}{\bibinfo{person}{Ronak Pradeep}, \bibinfo{person}{Nandan
  Thakur}, \bibinfo{person}{Sahel Sharifymoghaddam}, \bibinfo{person}{Eric
  Zhang}, \bibinfo{person}{Ryan Nguyen}, \bibinfo{person}{Daniel Campos},
  \bibinfo{person}{Nick Craswell}, {and} \bibinfo{person}{Jimmy Lin}.}
  \bibinfo{year}{2024}\natexlab{}.
\newblock \bibinfo{title}{{Ragnarök}: A Reusable RAG Framework and Baselines
  for TREC 2024 Retrieval-Augmented Generation Track}.
\newblock
\newblock
\showeprint[arxiv]{2406.16828}~[cs.IR]
\urldef\tempurl%
\url{https://arxiv.org/abs/2406.16828}
\showURL{%
\tempurl}


\bibitem[\protect\citeauthoryear{Qin, Jagerman, Hui, Zhuang, Wu, Yan, Shen,
  Liu, Liu, Metzler, Wang, and Bendersky}{Qin et~al\mbox{.}}{2024}]%
        {qin24:pairwise}
\bibfield{author}{\bibinfo{person}{Zhen Qin}, \bibinfo{person}{Rolf Jagerman},
  \bibinfo{person}{Kai Hui}, \bibinfo{person}{Honglei Zhuang},
  \bibinfo{person}{Junru Wu}, \bibinfo{person}{Le Yan},
  \bibinfo{person}{Jiaming Shen}, \bibinfo{person}{Tianqi Liu},
  \bibinfo{person}{Jialu Liu}, \bibinfo{person}{Donald Metzler},
  \bibinfo{person}{Xuanhui Wang}, {and} \bibinfo{person}{Michael Bendersky}.}
  \bibinfo{year}{2024}\natexlab{}.
\newblock \bibinfo{title}{Large Language Models are Effective Text Rankers with
  Pairwise Ranking Prompting}.
\newblock
\newblock
\showeprint[arxiv]{2306.17563}~[cs.IR]
\urldef\tempurl%
\url{https://arxiv.org/abs/2306.17563}
\showURL{%
\tempurl}


\bibitem[\protect\citeauthoryear{Reddy, Doo, Xu, Sultan, Swain, Sil, and
  Ji}{Reddy et~al\mbox{.}}{2024}]%
        {reddy24:first}
\bibfield{author}{\bibinfo{person}{Revanth~Gangi Reddy},
  \bibinfo{person}{JaeHyeok Doo}, \bibinfo{person}{Yifei Xu},
  \bibinfo{person}{Md~Arafat Sultan}, \bibinfo{person}{Deevya Swain},
  \bibinfo{person}{Avirup Sil}, {and} \bibinfo{person}{Heng Ji}.}
  \bibinfo{year}{2024}\natexlab{}.
\newblock \bibinfo{title}{FIRST: Faster Improved Listwise Reranking with Single
  Token Decoding}.
\newblock
\newblock
\showeprint[arxiv]{2406.15657}~[cs.IR]
\urldef\tempurl%
\url{https://arxiv.org/abs/2406.15657}
\showURL{%
\tempurl}


\bibitem[\protect\citeauthoryear{Robertson and Zaragoza}{Robertson and
  Zaragoza}{2009}]%
        {robertson09:bm25}
\bibfield{author}{\bibinfo{person}{Stephen Robertson} {and}
  \bibinfo{person}{Hugo Zaragoza}.} \bibinfo{year}{2009}\natexlab{}.
\newblock \showarticletitle{The Probabilistic Relevance Framework: BM25 and
  Beyond}.
\newblock \bibinfo{journal}{\emph{Foundations and Trends in Information
  Retrieval}} \bibinfo{volume}{3}, \bibinfo{number}{4} (\bibinfo{year}{2009}),
  \bibinfo{pages}{333--389}.
\newblock


\bibitem[\protect\citeauthoryear{Sun, Yan, Ma, Wang, Ren, Chen, Yin, and
  Ren}{Sun et~al\mbox{.}}{2023}]%
        {sun23:rankgpt}
\bibfield{author}{\bibinfo{person}{Weiwei Sun}, \bibinfo{person}{Lingyong Yan},
  \bibinfo{person}{Xinyu Ma}, \bibinfo{person}{Shuaiqiang Wang},
  \bibinfo{person}{Pengjie Ren}, \bibinfo{person}{Zhumin Chen},
  \bibinfo{person}{Dawei Yin}, {and} \bibinfo{person}{Zhaochun Ren}.}
  \bibinfo{year}{2023}\natexlab{}.
\newblock \bibinfo{title}{Is ChatGPT Good at Search? Investigating Large
  Language Models as Re-Ranking Agents}.
\newblock
\newblock
\showeprint[arxiv]{2304.09542}~[cs.CL]
\urldef\tempurl%
\url{https://arxiv.org/abs/2304.09542}
\showURL{%
\tempurl}


\bibitem[\protect\citeauthoryear{Tamber, Pradeep, and Lin}{Tamber
  et~al\mbox{.}}{2023}]%
        {tamber2023scalingdownlittingup}
\bibfield{author}{\bibinfo{person}{Manveer~Singh Tamber},
  \bibinfo{person}{Ronak Pradeep}, {and} \bibinfo{person}{Jimmy Lin}.}
  \bibinfo{year}{2023}\natexlab{}.
\newblock \bibinfo{title}{Scaling Down, LiTting Up: Efficient Zero-Shot
  Listwise Reranking with Seq2seq Encoder-Decoder Models}.
\newblock
\newblock
\showeprint[arxiv]{2312.16098}~[cs.IR]
\urldef\tempurl%
\url{https://arxiv.org/abs/2312.16098}
\showURL{%
\tempurl}


\bibitem[\protect\citeauthoryear{Thakur, Reimers, Rücklé, Srivastava, and
  Gurevych}{Thakur et~al\mbox{.}}{2021}]%
        {thakur21:beir}
\bibfield{author}{\bibinfo{person}{Nandan Thakur}, \bibinfo{person}{Nils
  Reimers}, \bibinfo{person}{Andreas Rücklé}, \bibinfo{person}{Abhishek
  Srivastava}, {and} \bibinfo{person}{Iryna Gurevych}.}
  \bibinfo{year}{2021}\natexlab{}.
\newblock \bibinfo{title}{BEIR: A Heterogenous Benchmark for Zero-shot
  Evaluation of Information Retrieval Models}.
\newblock
\newblock
\showeprint[arxiv]{2104.08663}~[cs.IR]
\urldef\tempurl%
\url{https://arxiv.org/abs/2104.08663}
\showURL{%
\tempurl}


\bibitem[\protect\citeauthoryear{Tunstall, Beeching, Lambert, Rajani, Rasul,
  Belkada, Huang, von Werra, Fourrier, Habib, Sarrazin, Sanseviero, Rush, and
  Wolf}{Tunstall et~al\mbox{.}}{2023}]%
        {tunstall23:zephyr}
\bibfield{author}{\bibinfo{person}{Lewis Tunstall}, \bibinfo{person}{Edward
  Beeching}, \bibinfo{person}{Nathan Lambert}, \bibinfo{person}{Nazneen
  Rajani}, \bibinfo{person}{Kashif Rasul}, \bibinfo{person}{Younes Belkada},
  \bibinfo{person}{Shengyi Huang}, \bibinfo{person}{Leandro von Werra},
  \bibinfo{person}{Clémentine Fourrier}, \bibinfo{person}{Nathan Habib},
  \bibinfo{person}{Nathan Sarrazin}, \bibinfo{person}{Omar Sanseviero},
  \bibinfo{person}{Alexander~M. Rush}, {and} \bibinfo{person}{Thomas Wolf}.}
  \bibinfo{year}{2023}\natexlab{}.
\newblock \bibinfo{title}{Zephyr: Direct Distillation of LM Alignment}.
\newblock
\newblock
\showeprint[arxiv]{2310.16944}~[cs.LG]
\urldef\tempurl%
\url{https://arxiv.org/abs/2310.16944}
\showURL{%
\tempurl}


\bibitem[\protect\citeauthoryear{Zhang, {Hofst\"{a}tter}, Lewis, Tang, and
  Lin}{Zhang et~al\mbox{.}}{2023}]%
        {zhang23:rankwogpt}
\bibfield{author}{\bibinfo{person}{Xinyu Zhang}, \bibinfo{person}{Sebastian
  {Hofst\"{a}tter}}, \bibinfo{person}{Patrick Lewis}, \bibinfo{person}{Raphael
  Tang}, {and} \bibinfo{person}{Jimmy Lin}.} \bibinfo{year}{2023}\natexlab{}.
\newblock \bibinfo{title}{{Rank-without-GPT}: Building {GPT}-Independent
  Listwise Rerankers on Open-Source Large Language Models}.
\newblock
\newblock
\showeprint[arxiv]{2312.02969}~[cs.CL]
\urldef\tempurl%
\url{https://arxiv.org/abs/2312.02969}
\showURL{%
\tempurl}


\bibitem[\protect\citeauthoryear{Zhu, Yuan, Wang, Liu, Liu, Deng, Chen, Liu,
  Dou, and Wen}{Zhu et~al\mbox{.}}{2024}]%
        {zhu24:llm_ir_survey}
\bibfield{author}{\bibinfo{person}{Yutao Zhu}, \bibinfo{person}{Huaying Yuan},
  \bibinfo{person}{Shuting Wang}, \bibinfo{person}{Jiongnan Liu},
  \bibinfo{person}{Wenhan Liu}, \bibinfo{person}{Chenlong Deng},
  \bibinfo{person}{Haonan Chen}, \bibinfo{person}{Zheng Liu},
  \bibinfo{person}{Zhicheng Dou}, {and} \bibinfo{person}{Ji-Rong Wen}.}
  \bibinfo{year}{2024}\natexlab{}.
\newblock \bibinfo{title}{Large Language Models for Information Retrieval: A
  Survey}.
\newblock
\newblock
\showeprint[arxiv]{2308.07107}~[cs.CL]
\urldef\tempurl%
\url{https://arxiv.org/abs/2308.07107}
\showURL{%
\tempurl}


\bibitem[\protect\citeauthoryear{Zhuang, Qin, Hui, Wu, Yan, Wang, and
  Bendersky}{Zhuang et~al\mbox{.}}{2024a}]%
        {zhuang24:binary}
\bibfield{author}{\bibinfo{person}{Honglei Zhuang}, \bibinfo{person}{Zhen Qin},
  \bibinfo{person}{Kai Hui}, \bibinfo{person}{Junru Wu}, \bibinfo{person}{Le
  Yan}, \bibinfo{person}{Xuanhui Wang}, {and} \bibinfo{person}{Michael
  Bendersky}.} \bibinfo{year}{2024}\natexlab{a}.
\newblock \bibinfo{title}{Beyond Yes and No: Improving Zero-Shot LLM Rankers
  via Scoring Fine-Grained Relevance Labels}.
\newblock
\newblock
\showeprint[arxiv]{2310.14122}~[cs.IR]
\urldef\tempurl%
\url{https://arxiv.org/abs/2310.14122}
\showURL{%
\tempurl}


\bibitem[\protect\citeauthoryear{Zhuang, Zhuang, Koopman, and Zuccon}{Zhuang
  et~al\mbox{.}}{2024b}]%
        {zhuang24:setwise}
\bibfield{author}{\bibinfo{person}{Shengyao Zhuang}, \bibinfo{person}{Honglei
  Zhuang}, \bibinfo{person}{Bevan Koopman}, {and} \bibinfo{person}{Guido
  Zuccon}.} \bibinfo{year}{2024}\natexlab{b}.
\newblock \bibinfo{title}{A Setwise Approach for Effective and Highly Efficient
  Zero-shot Ranking with Large Language Models}.
\newblock
\newblock
\urldef\tempurl%
\url{https://doi.org/10.1145/3626772.3657813}
\showDOI{\tempurl}
\showeprint[arxiv]{2310.09497}~[cs.IR]


\end{thebibliography}
